# Preserving Patient-centred Controls in Electronic Health Record Systems: A Reliance-based Model Implication

Pasupathy Vimalachandran[1], Hua Wang [2], Yanchun Zhang [2], Ben Heyward [3]
and Yueai Zhao [4]

[1] *Centre for Applied Informatics, College of Engineering and Science,*
*Victoria University, Melbourne, Australia*
*Pasupathy.Vimalachandran@live.vu.edu.au*

[2] *Centre for Applied Informatics, College of Engineering and Science,*
*Victoria University, Melbourne, Australia*
*{hua.wang,yanchun.zhang}@vu.edu.au*

[3] *Nexus Online, Adelaide, Australia. benhey37@gmail..com*

[4] *Taiyuan Normal University, Shanxi Province, China. tysyzya@sina.com*

**Abstract-** *As a consequence of the huge advancement of the Electronic Health Record (EHR) in healthcare settings, the My Health Record (MHR) is introduced in Australia. However security and privacy of the MHR system have been encumbering the development of the system. Even though the MHR system is claimed as patient-centred and patient-controlled, there are several instances where healthcare providers (other than the usual provider) and system operators who maintain the system can easily access the system and these unauthorised accesses can lead to a breach of the privacy of the patients. This is one of the main concerns of the consumers that affect the uptake of the system. In this paper, we propose a patient centred MHR framework which requests authorisation from the patient to access their sensitive health information. The proposed model increases the involvement and satisfaction of the patients in their healthcare and also suggests mobile security system to give an online permission to access the MHR system.*

## I. Introduction

Patient centred care is the practice of caring for patients (and their families) in ways that are meaningful and valuable to the individual patient. It includes listening to, informing and involving patients in their care. The American Institute of Medicine (IOM) defines patient-centred care as: "Providing care that is respectful of, and responsive to, individual patient preferences, needs and values, and ensuring that patient values guide all clinical decisions" [8]. Picker Institute and Harvard Medical School has defined the Patient-Centred care through eight principles [9] which are explained in Figure 1.

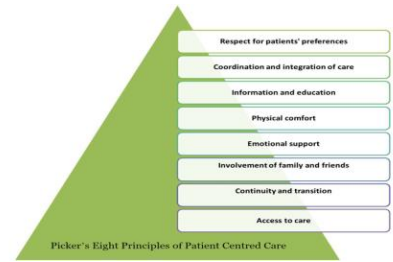

*Figure 1: Picker's Eight Principles of Patient Centred Care [Source [9]]*

In the Picker's patient-centred principles, more than one principle directly linked with patients' privacy and security (e.g. Respect for patients' preferences, information and education and involvement of family and friends). The researchers also found that there are certain practices conducive to a positive patient experience and their findings form Picker's Eight Principles of Patient-Centred Care. Patient-centred care is a quality of personal, professional and organisational relationships. Thus, efforts to promote patient-centred care should consider patient-centredness of patients (and their families), clinicians and health systems [10, 11].

We review the previous works in Section 2. In Section 3, we propose our new method to preserve privacy and security for the MHR system in Australia. The implementation of the proposed model including computer program, sequence diagram and the communication method are discussed in Section 4. Section 5 analyses possible concerns and solutions for those concerns of the proposed model. The paper concludes and leaves future development suggestions in Section 6.

## 2. Related work

Several resolutions have been proposed to overcome the privacy and security related issues with EHR. Most of the resolutions relate to access control and/or cryptographic approaches.

Cyptography methods are also considered as the safest way approach to preserve the privacy and security of cloud based systems including EHR. To transmit the data safely in cloud computing, cryptography solutions are suitable enough by practicing the public key structure (27,39,41). The aim of the cryptography is to encrypt the confidential private information including clinical related details before sending and storing in the cloud. However, in the practical world, this is not the case all the time. The system operator has more power and the patients' ability is very limited in these settings. Understanding these limitations, there were some techniques proposed. Benaloh Chase et al., Jin Ahn et al. and Li et al. recommend models that patients encrypt their health information before sending and storing in the cloud to overcome the potential risks around the privacy exposure with system operators [28, 29, 30,37].

Van der Haak et al. use digital signatures and PKI (public key infrastructure) authentications to satisfy legal requirements for the exchange of the EMR [31]. Pseudonymation techniques used to preserve patients anonymity by Ateniese at al. [32]. A health communication monitoring method to observer the health information exchange is also proposed by Layouni et al. [33]. Even though using the cryptography approaches is safe while storing the sensitive health information, accessing that critical information when it is required seems a difficult task. For that reason, the technique decays the whole idea and purpose of the cryptography approaches in EHR settings as accessing the right health information in the right time at the right place is the key of the EHR systems [34,43,44.45].

### 3. The Proposed Model

In the proposed model, healthcare provider normal requests will require patients' permission to access their EHR. The requests from healthcare providers can be categorised into two main groups; special request and normal request. Life threatening circumstances are special request and daily routine activities such as follow ups and recalls and reminders all fallen into normal request. The normal requests to access patients' EHR for these non-life threatening or non- emergency situations required the patients' permission.

The patients' permission system really works like an online consent method and includes authentication and authorisation processes.

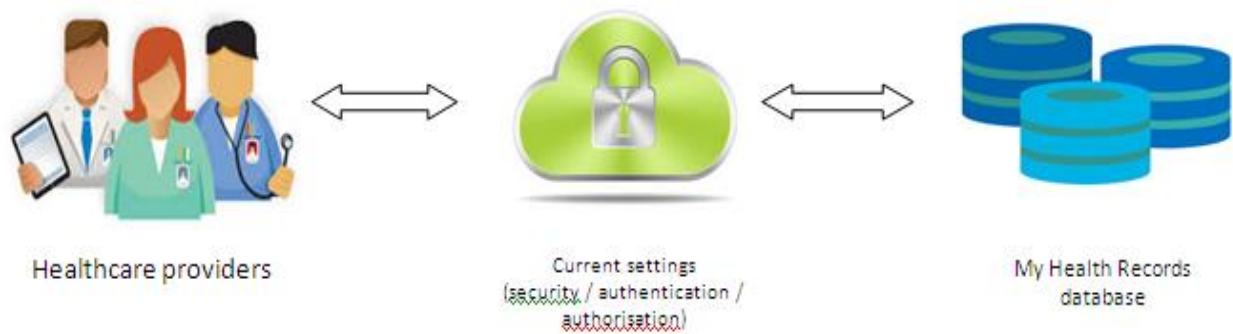

*Figure 2: Existing EHR model*

Current settings may include the following servers:
(i) Authentication Server: The authentication server uses to authenticate the credentials, usually usernames and passwords of the EHR users. When a user submits a valid set of credentials, it receives a cryptographic ticket that it can subsequently use to access various services.

(ii) Access Control List (ACL) Server: This server verifies the users when they provide their credentials; usernames and passwords. ACL is basically a table that tells the system which access rights each user has to a particular object, such as a patient record. Each object has a security attribute that identifies its ACL and specifies how a user can access a patient's EHR and perform actions. For example a radiology technician usually does not need to access a patient's mental health information in an EHR (read – action) or a health insurance company does not need to change a patient medical history in an EHR.
(iii) Authentication Server: Once a request passed through the authentication and ACL processes, the users must be authorised by the system to access the actual patient EHR.

However, in our proposed model, the patient whose record is accessed would be the key of the system. When a user attempts to access a patient's EHR, the user requires the patient's acceptance to proceed further. This model will not only preserve the patient centric concept but also prevent misuse of the EHR system. To access sensitive health information through this model healthcare providers require the permission from the actual patient. This means, every healthcare provider is made aware that when they login the EHR system, the patient is notified and followed. The patients can keep track of why and what is being accessed and what changes have been made in their record. This concept will provide a high security and privacy environment for EHRs.

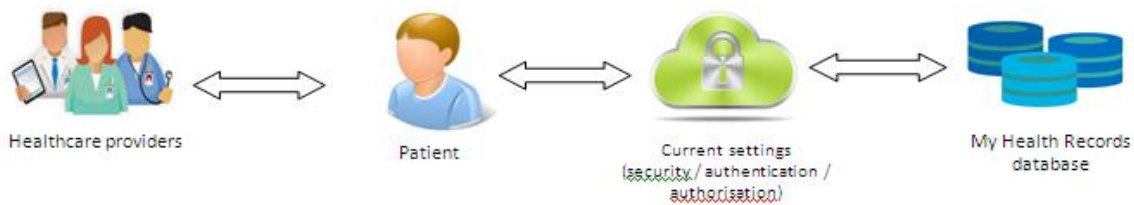

*Figure 3: The proposed EHR model*

The next challenge in this model is to ensure effective communication between the patients and the healthcare providers when a healthcare provider requires access to their usual patients. Mobile devices are the most feasible option to perform this communication. Deloitte's Mobile Consumer Survey 2015 [35] reveals that a massive 79% of Australian now own a smart phone, with an even higher penetration rate expected next year which means 2016.

**4. Development and evaluation of the proposed model**

In the proposed model, the user is generally a healthcare providers or their organisation that needs to access their patients' MHR. A user can be a GP, medical specialist, allied health professional or pharmacists. The purpose of accessing the MHR may differ from user to user. While a GP accesses patients' previous records to write a prescription, a Pharmacist accesses to verify the medication that prescribed by a GP. Authentication Server, ACL Server and Authorisation Server are discussed in Section 3. Patient 'A' is the actual patient who has a MHR and will give access permission for any healthcare providers who requests access. The healthcare provider cannot be the usual GP as the usual GP access permission is provided as default by the proposed system. The EHR database is the system that handles millions of patients' MHR. The EHR database contains high capacity storage devices and controls and has got capacity to increase storage and process power.

'Control System' is the device that reflects for our proposed system. New rules can be added and removed with the system. The Control System includes a mobile security system that performs the following tasks:
(i) receive an access requests from a user
(ii) verify the right method for the patient
(iii) send the access request to the patient
(iv) receive the access response from the patient
(v) send the access response to the Authorisation Server to react according to the response.

An instance of the health information access method of the MHR is illustrated below in Figure 4.

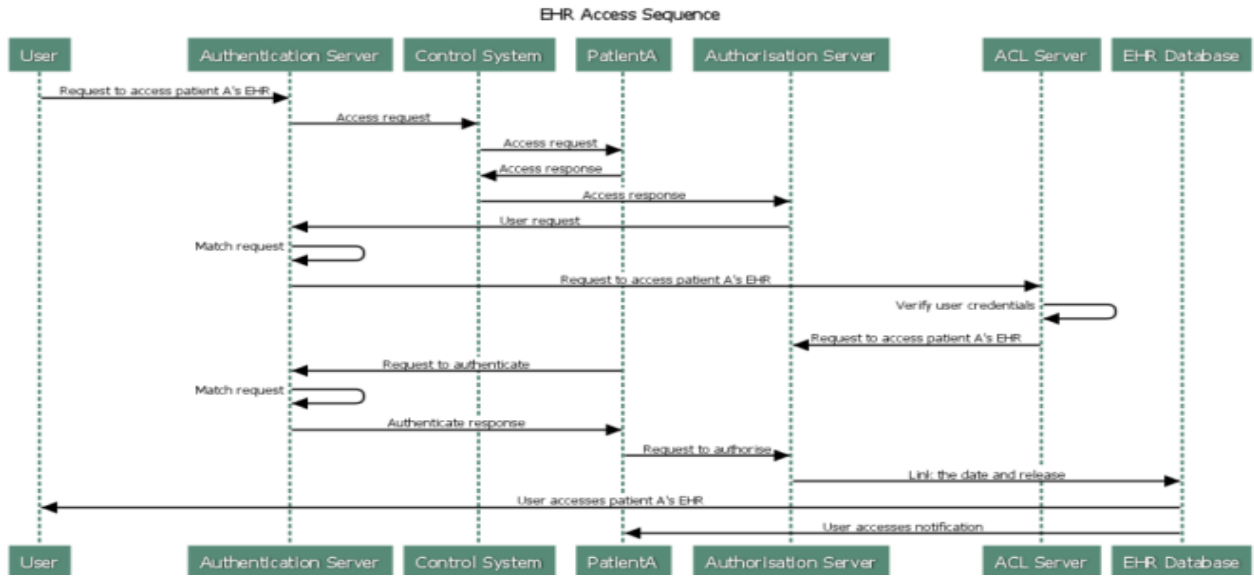

*Figure 4: An instance of data accessing method from EHR using a sequence diagram*

To develop and evaluate, the following computer programming code has been used. First of all every healthcare provider is assigned to get permission from their usual patients.

```
Function checkdata() As Boolean
bcheck = True
   If txtuser.Text = "" Then
MsgBox "Enter user name", , "HighSec System"
txtuser.SetFocus
bcheck = False
      Exit Function
   End If
   If txtpassword.Text = "" Then
MsgBox "Enter Password", , "HighSec System"
txtpassword.SetFocus
bcheck = False
      Exit Function
   End If
    If cmbUsertype.Text = "Normal" Then
      If cmbManager.Text = "Select" Then
MsgBox "Select Manager", , "HighSec System"
txtpassword.SetFocus
bcheck = False
        Exit Function
      End If
    End If
checkdata = bcheck
End Function

Private Sub cmbUsertype_Click()
   If cmbUsertype.Text = "Normal" Then
      Frame1.Visible = True
   Else
      Frame1.Visible = False
   End If
End Sub
```

*Code 1 :The patient assignment*

```
Private Sub cmdok_Click()
bcheck = checkdata
usertype = Left(cmbUsertype.Text, 1)
 If bcheck = checkdata Then
rs.Open "select * from usertable where userid='"
+ txtuser.Text + "'", cn
     If Not rs.EOF And Not rs.BOFThen
MsgBox "This user already exists", , "HighSec
System"
     Else
newpwd = encryptdata(txtpassword.Text,
newkey)

newpwd = txtpassword.Text
ssql = "insert into usertable
(userid,pwd,usertype) values('" + txtuser.Text +
"','" + newpwd + "','" + usertype + "')"
InputBox "", ,ssql

cn.Executessql
       If usertype = "N" Then
ssql = "insert into groupuser (user1,user2)
values('" + txtuser.Text + "','" +
cmbManager.Text + "')"
cn.Executessql
       End If
ans = MsgBox("User created succesfully. " +
vbCrLf + " Do you want to close this window?",
vbYesNo)
       If ans = vbYes Then
          Unload Me
       Else
txtuser.Text = ""
txtpassword.Text = ""
       End If
     End If
rs.Close
```

```
        End If
End Sub

Private Sub Form_Load()
ssql = "select * from usertable where
usertype='M'"
rs.Openssql, cn
While Notrs.EOF
cmbManager.AddItemrs(0)
rs.MoveNext
Wend
rs.Close
cmbUsertype.ListIndex = 0
End Sub
```

*Code 2: Creating the patient for healthcare provider*

Once the patient is determined for a healthcare provider to get the permission, the message must be sent out through an effective communication line to and from as quick as possible. This speedy process can be completed using a mobile security system. A mobile security system will be a best solution for this proposed model. Once the mobile security two-factor authentication is enrolled the healthcare provider can login as usual with their username and password to access the EHR system and then the patient will receive a message through the device for the permission. This process can be done via SMS, voice call, one-time pass code or the mobile smart phone apps. For example, 'Duo' mobile security system [36] has its own smart phone app to do the two-factor authentication verification process. The system also lets patients link multiple devices to the account such as mobile phone and a landline, a landline and hardware token or two different mobile devices [36]. This will provide increased accessibility for healthcare provider. As mentioned in Figure 5 below, the mobile security system will provide an additional security layer for the EHR systems. The username and password that has been created for the system remain the same and once provided, the patients' approval request will be sent out to the patients preferred communication device/s. Consequently, if a healthcare provider wants to access patients' sensitive health information for any reason (e.g. follow up or targeting high chronic disease patients to send a reminder), then an authorised person (actual patient) must give permission.

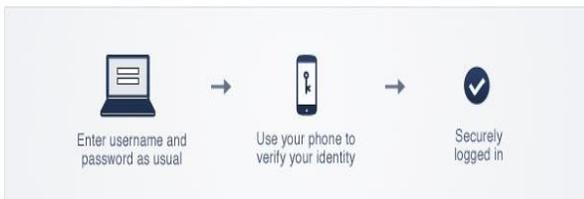

*Figure 5: How a Mobile Security System Works (source: [36])*

When creating user login for a healthcare provider to access past medical history using a clinical software, a patient link must be established as illustrated through above computer program coding (Code 2). Moreover a patient has more than one option to approve or deny the healthcare provider login request. For example, following options are available with Duo mobile security system.

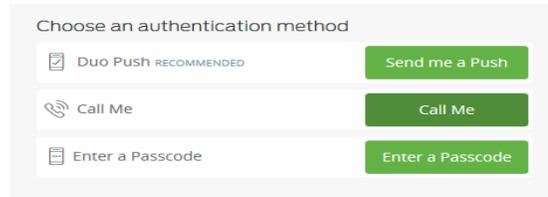

*Figure 6: Authentication Methods for mobile security system*

Figure 6 above illustrates three various options available with Duo mobile security systems for a healthcare provider to receive a response from a patient.

**5. Analysis of the proposed model**

From a patient point of view, the proposed model provides an additional privacy and security option. A majority of the patients' group who use smart phones would like this model and the idea helps to increase their level of trust on EHR systems in the future. However from a healthcare provider perspective it may be another step to go through to complete their routine activities and even more time consuming in some cases.

With any model, satisfying both, patients and healthcare providers would be an ongoing challenge. For example our proposed model seems to be thinking about the patient privacy, than healthcare provider accessing health information in time. However, we hope our proposed model will improve with the following;
- after mobile coverage black spot issue has been resolved,
- with increase use of mobile devices and
- after the completion of the IBN (Internet Broadband Network) program.

In other words, even though the proposed model has some challenges in its introduction, there will be more possibilities to accept by healthcare providers in the future. The problems that identified are discussed in Table 1. The Table 1 also gives solutions for those problems.

| Problem | Description / possible solution |
|---|---|
| If any patient not using smart phone | In Australia, 79% of population use smart phones. Any elderly or younger patients can nominate someone who has a mobile device to give permission. |
| If a patient is on holidays and has no access to any mobile device | When a patient goes on holiday or is not accessible via mobile device they can assign someone who can access for a short time period on their behalf to receive the message |
| More time consuming option for a healthcare provider | A patient can assign more than one mobile devices to receive the message |
| Under emergency and life threatening circumstances, a healthcare provider cannot wait to receive a patient permission. | A healthcare provider can go the emergency five-day permission free option, but the healthcare provider must be responsible for the access. The detail of the access will be sent out to the patient by email. |
| The actual patient who needs to give permission is not in a position to give it e.g. the patient meets an accident | A healthcare provider can go to the emergency five-day permission free option, but the healthcare provider must be responsible for the access. The detail of the access will be sent out to the patient by email. |

*Table 1: Problems and solutions for the proposed model*

The other consideration is why and how frequently the healthcare providers or their organisations access a patient EHR while the patient is not present (i.e. except consultation time). In a healthcare provider organisation setting and the arrangements, there are many instances when a staff or healthcare provider accesses a patient's health information for many purposes including the followings:
- to monitor health conditions
- to review health information
- to review pathology and radiology results
- to remind and recall procedures
- to scan and store test results
- to target a high risk patient including chronic disease.

For high risk category patients, the healthcare provider organisations need to access their health information very frequently in order to deliver a quality, well-informed and continuous healthcare.

**6. Conclusion and future suggestions**

In this paper, we present a new authorisation concept to preserve the privacy and security of the health information in the MHR. In EHR systems, the patients' knowledge of what is happening once the health information is uploaded to MHR is lacking. With the Patient-Centred health concept this lacking knowledge must be satisfied. Our proposed model satisfies and keeps the patient at the central point in providing health care for them. We also provide a solution in communication between patients and the healthcare providers using mobile security method. The development process is also discussed using a computer programming language code in assigning the right patient for the healthcare provider.

To examine the overall changes and communication, a monitoring and tracking system may be needed to keep the preserve in high level. This system will give an additional security layer for the proposed model. Proposing a method to include non-mobile device users to give authentication might be another area for future development.